\input epsf

\documentstyle[12pt]{article}

\thicklines                         
\def\a{\alpha}
\def\b{\beta}

\def\d{\delta}
\def\e{\epsilon}                
\def\f{\phi}                    

\def\h{\eta}

\def\j{\psi}

\def\l{\lambda}
\def\m{\mu}
\def\n{\nu}
\def\o{\omega}
\def\r{\rho}                    
\def\s{\sigma}                  

\def\x{\xi}

\def\F{\Phi}
\def\G{\Gamma}

\def\cbo{{\,\raise-.15ex\Sc [\,}}                       
\def\dg{^\dagger}                                     

\def\svev#1{\left\langle #1\right\rangle}       

\def\ddt#1{{\buildrel {\hbox{\LARGE .\kern-2pt.}} \over {#1}}}

\def\beqn#1{ \renewcommand{\theequation}{#1}
             \begin{eqnarray} }
\def\eeqn{ \renewcommand{\theequation}{\arabic{equation}}
           \end{eqnarray} }

\def\beqr#1{ \setcounter{equation}{#1}
             \begin{eqnarray} }

\def\eeqr{\end{eqnarray}}
\def\NON{\nonumber\\}
\def\beqrabc#1{ \setcounter{equation}{0}
                \renewcommand{\theequation}{#1\alph{equation}}
                \begin{eqnarray} }
\def\beqrn#1#2{ \setcounter{equation}{#2}
                \renewcommand{\theequation}{#1.\arabic{equation}}
                \begin{eqnarray} }
\def\seeq#1{eq.~(\ref{#1})}
\def\seEq#1{Eq.~(\ref{#1})}
\def\seeqs#1{eqs.~(\ref{#1})}

\def\seneq#1{~(\ref{#1})}

\def\NPB#1{Nucl. Phys. {\bf B#1}}

\def\PLB#1{Phys. Lett. {\bf B#1}}
\def\PRD#1{Phys. Rev. {\bf D#1}}

\def\PRL#1{Phys. Rev. Lett. {\bf #1}}
\def\PRP#1{Phys. Rep. {\bf #1}}

\def\sstyle{\scriptstyle}

\def\ie{\mbox{i.e.} }
\def\eg{\mbox{e.g.} }

\def\frac#1#2{ {\sstyle {#1\over #2} } }

\def\tr{{\rm tr}\,}

\def\half{{1\over 2}}
\def\Re{{\rm Re\,}}

\def\beq{\renewcommand{\theequation}{\arabic{equation}}
         \begin{equation}}
\def\eeq{\end{equation}}
\def\bqry{\renewcommand{\theequation}{\arabic{equation}}
          \begin{eqnarray}}
\def\eqry{\end{eqnarray}}

\renewcommand{\theequation}{\arabic{equation}}

\renewcommand{\thefootnote}{\fnsymbol{footnote}}

\begin{document}
\hyphenation{fer-mio-nic per-tur-ba-tive}

\noindent  \hfill TAUP--2222--94

\par
\begin{center}
\vspace{15mm}
{\large\bf
  Explicit Breaking of Supersymmetry by\\
  Non-Perturbative Effects}
\\[10mm]
A.\ Casher, V.\ Elkonin and Y.\ Shamir\footnote{
  Work supported in part by the US-Israel Binational Science Foundation,
  and the Israel Academy of Science.}\\[5mm]
School of Physics and Astronomy\\
Beverly and Raymond Sackler Faculty of Exact Sciences\\
Tel-Aviv University, Ramat Aviv 69978, ISRAEL\\[15mm]
{ABSTRACT}\\[2mm]
  \end{center}
\begin{quotation}

   Instanton effects in a family of completely massive Higgs models with
N=1 supersymmetry are investigated. The models have $N_c=2$ and
\mbox{$N_f\ge 2$}.
In each model, we show that a certain gauge invariant correlation function
depends in a non-trivial way on its coordinates, in spite of the fact that
supersymmetry requires its constancy. This means that non-perturbative
effects break supersymmetry explicitly in the one instanton sector.
We also show that condensates arising in the point-like limit
of the above correlation functions can in principle be used to induce
the Electro-Weak scale.

\end{quotation}

\setcounter{footnote}{0}
\renewcommand{\thefootnote}{\arabic{footnote}}

\newpage
\noindent {\large\bf 1.~~Introduction}
\vspace{3ex}

  Non-perturbative effects in asymptotically free, supersymmetric gauge
theories have been investigated extensively for more than a decade~[1-10].
Recently there has been a renewed interest in the subject~\cite{sw}.

  Crucial to almost the entire literature on the subject, is the
{\it assumption} that non-perturbative effects do not break supersymmetry
(SUSY) explicitly. This applies in particular to the study of dynamical
(spontaneous) SUSY breaking~[1-5]. There have been
attempts to verify the validity of that assumption~[4-8], but with no
conclusive results.

  In a continuum framework, the leading non-perturbative effect arises from
the one instanton sector. For a quantity that vanishes to all orders in
perturbation theory, the one instanton result represents the leading order
contribution in a systematic expansion {\it provided} the gauge coupling is
{\it weak}. For this reason, we restrict our attention in this paper to
SUSY-Higgs models.

  We calculate various gauge invariant
correlation functions in a family of SU(2)-Higgs models with N=1 SUSY.
The computation involves little more than semi-classical instanton calculus,
and it reveals the existence of explicit SUSY breaking effects in the one
instanton sector.

  In each model, the classical potential has a unique
supersymmetric minimum. The Higgs VEV breaks the SU(2) gauge symmetry
completely, and all fields acquire non-zero masses at tree level.
Because of the absence of massless fermions there is no room for spontaneous
SUSY breaking, as there is no candidate to become a goldstino\footnote{
  Since the broken gauge coupling can be taken to be as weak as we like,
  we can safely assume that there are no composite massless states.}.
We show that, nevertheless,  a certain gauge invariant correlation function
violates a SUSY Ward identity. This proves that SUSY is broken explicitly in
the one instanton sector.  The Ward identity has been chosen to
minimize the amount of technicalities involved in the computation.
The simplest model has $N_f=2$ in the terminology of ref.~\cite{ads},
and in App.~B we generalize our results to $N_f > 2$.

  In more detail, we first show that a certain bosonic condensate is formed.
The operator which condenses takes the form $\G(x,x)$, where $\G(x,y)$
is the correlator of two gauge invariant composite operators. We then show
that $\G(x,y)\to 0$ as $\mbox{$|x-y|$}\to\infty$.
The discrepancy with SUSY arises
because a SUSY Ward identity requires the correlator $\G(x,y)$ to be
independent
of the separation $x-y$. We expect that the computation of other quantities
of physical interest, such as non-perturbative corrections to boson and
fermion masses, will reveal further violations of SUSY.

  This paper is organized as follows. In Sect.~2 we define the basic SUSY-Higgs
model.  In Sect.~3 we discuss the one instanton sector and find that a certain
bosonic condensate is formed. In Sect.~4 we show that SUSY is explicitly broken
in the one instanton sector. In Sect.~5 we make  a first excursion into the
phenomenological implications of our result.  We show that condensates of the
kind described above can in principle be used to induce the Electro-Weak scale.
Sect.~6 contains a short discussion.
In App.~A we discuss the asymptotic behaviour of the zero modes. Finally, in
App.~B we show that violations of SUSY occur in models with arbitrary
$N_f\ge 2$.

\vspace{5ex}
\noindent {\large\bf 2.~~The model}
\vspace{3ex}

  The model we present in this section is a variant of one of the simplest SUSY
Higgs models. This model serves as a test case.  We show that a certain SUSY
Ward identity pertaining to a gauge invariant correlation function is violated
explicitly. Our results also have phenomenological implications.  We find that
certain bosonic condensates are formed quite generally in the one instanton
sector.  If these condensates arise from GUT scale physics or, alternatively,
from a strongly interacting hidden sector at the TeV range, they can in
principle be used to induce the Electro-Weak scale in the observed sector.

  The basic ingredient of the model is an SU(2)-Higgs sector. It consists
of a gauge supermultiplet $(A^a_\m,\l^a)$ and of several chiral
supermultiplets. $\F_{iA}=(\f_{iA},\j_{iA})$ contain the Higgs fields and
their fermionic partners. In addition, there is a neutral supermultiplet
$\F^0=(\f^0,\j^0)$.
The higgs fields are doublets of both SU(2)$_c$ and SU(2)$_f$,
and $A,i=1,2$ are respectively the colour and flavour indices. We let $T^a$
and $F^a$ denote the colour and flavour generators respectively.
In this paper, we use the representation
$T^a_{AB} = -\half \s^a_{BA}$ and $F^a_{ij} = \half \s^a_{ij}$.

  The superpotential is
\beq
  W_1 = h \F^0 \left(
       {1\over 2} \e_{ij} \e_{AB} \F_{iA} \F_{jB} - v^{2}
       \right)\,.
\label{sp}
\eeq
The classical potential has a unique (up to colour and flavour transformations)
SUSY minimum, which can be chosen to be
\beq
  \svev{\f_{iA}} = v \d_{iA} \,.
\eeq
The minimum breaks the gauge symmetry completely, and it leaves unbroken
the diagonal SU(2)$_V$ generated by $T^a + F^a$. Under SU(2)$_V$, the
two Higgs superfields decompose into a singlet
$\F' = \d_{iA} \F_{iA}/\sqrt{2}$
and a triplet $\F^a = \s^a_{iA} \F_{iA}/\sqrt{2}$.

  All fields acquire non-zero masses at tree level. The massive gauge
supermultiplet has mass $\m=gv$ where $g$ is the gauge coupling.
Its bosonic and fermionic components are
respectively $A^a_\m$ and $\Re \f^a$, and  $\l^a$ and $\bar\j^a$.
(The fields  $(\l^a, \bar\j^a)$ form a massive Dirac spinor).
In the singlet sector, the mass is $m=\sqrt{2}hv$, and the component fields
are $\f^0$, $\f'$, $\j^0$ and $\bar\j'$.

  We add to the model two ``lepton'' families. The
corresponding chiral superfields are $\h_A^\pm$ and  $\x_i^\pm$.
These letters will also be used to denote the fermionic components, whereas a
tilde over the letter is used to denote the
scalar components. The $\pm$ superscript actually correspond to a new SU(2)
``family'' symmetry. Apart from the fact that there are two $\h$-s and
two $\x$-s, the family SU(2) will play little role below.

  The full superpotential is $W=W_1+W_2$, where
\beq
  W_2 = y\, \e_{ij} \e_{AB}\, \F_{jB}
        \left( \x^+_i\, \h^-_A - \x^-_i\, \h^+_A  \right) \,.
\label{sp2}
\eeq
The two ``lepton'' families are massive too, and their mass
is $m_1 = yv$. The Dirac spinors are $(\h^\pm, \bar\x^\mp)$.
A summary of the field content of the model can be found in Table~1.
This table also gives the charges of the fermions under the non-anomalous
$R$-symmetry. For chiral superfields, the charges of the corresponding
scalars are related by $Q_R(scalar) = Q_R(fermion)+1$.

\vspace{5ex}
\noindent {\large\bf 3.~~The one instanton sector}
\vspace{3ex}

  There are standard techniques to compute correlation functions
in the one instanton sector of any Higgs model. One integrates over a
family of classical backgrounds labeled by collective coordinates, and
for every background one has a systematic expansion in powers of the
coupling constant(s). In a SUSY-Higgs model there are exact fermionic
zero modes in spite of the fact that some (or all) fields are massive.
This feature, however, is not unique to SUSY theories, and it is present
already in Electro-Weak sector of the Standard Model.
The physical significance of the fermionic zero modes
and the techniques for dealing with them have been
discussed by 'tHooft~\cite{tft}.

  In this paper we follow the conventions of ref.~\cite{ads} with minor
modifications. The classical gauge field is given by
\beq
  A^a_\m = {2\over g}\, a(r)\, \bar\h_{a\m\n}\, (x^\m-x^\m_0) \,,
\label{a}
\eeq
where $r^2 = (x-x_0)^2$. The collective coordinates $x_0^\m$ describe the
instanton's center.  The function $a(r)$ tends to a non-zero constant at the
origin, and its asymptotic behaviour is $a(r)\sim 1/r^2$. In the case of an
unbroken gauge symmetry, one has $a(r)=1/(r^2 +\r^2)$ where $\r$ is the
instanton's size. In the Higgs case, the constrained instanton~\cite{constr}
is still characterized by a scale parameter $\r$, but the precise form of
$a(r)$ is different.

  The Higgs field has the following form
\beq
  \f_{iA} = i v\, \bar\s^\m_{iA}\, (x^\m-x^\m_0)\, \varphi(r) \,.
\label{higgs}
\eeq
The real function $\varphi(r)$ tends to a constant at small $r$,
whereas its asymptotic behaviour is $\varphi(r)\sim 1/r$. Finally,
the conserved angular momenta in the instanton background are
\bqry
   K^a_1 & = & S^a_1 + L^a_1 + T^a \,, \NON
   K^a_2 & = & S^a_2 + L^a_2 + F^a \,.
\eqry

  We now turn to the fermionic zero modes. In the absence of a Higgs VEV,
the model had had four gaugino zero modes and four matter zero modes (one for
each charged doublet). With the Higgs VEV, four of the zero modes
disappear~\cite{ads}. Two zero modes survive in the SU(2)-Higgs sector.
We refer to them as the ``gaugino'' zero modes.	In addition, every ``lepton''
family contributes one zero mode. Another feature is that,
in the Higgs case, each zero mode contains more than one channel.
The ``gaugino'' zero modes have the following decomposition (for $x_0^\m=0$)
\bqry
   (\l^a_\a)_k & = & \s^a_{\a k}\, f(r) \,, \NON
   (\j\dg_{iA\a})_k & = & i \d_{i\a} \d_{Ak}\, g(r)
                     +i x^\m x^\n \s^\m_{Ai} \bar\s^\n_{\a k}\, h(r) \,, \NON
   (\j^0_\a)_k & = & i \d_{\a k}\, p(r) \,.
\label{gg}
\eqry
Here $\a$ is the spinor index, and the index $k=1,2$ counts the two zero modes.
We use the notation $\j^\dagger_\a = \e_{\a\b} \bar\j_\b$.
The ``lepton'' zero modes are
\bqry
   \h^\pm_{\a A} & = &  \d_{\a A}\, u(r) \,, \NON
   \x^{\dagger\mp}_{\a i} & = & \mp \d_{\a i}\, v(r) \,.
\label{lept}
\eqry
The quantum numbers of the four zero modes as
well as their different channels can be found in Table~2.

  For each zero mode, the radial functions solve a set of ordinary coupled
differential equations. These equations can be found in App.~A, which also
gives the asymptotic large-$r$ behaviour of the zero modes.
The small-$r$ behaviour will not be needed below.

  In the rest of this section we will show that a certain bosonic condensate
is formed in the one instanton sector. In the next section we show that,
with slight modification, this condensate can be regarded as the point-like
limit of a gauge invariant correlation function, and that that correlation
function violates SUSY by failing to be a constant.

\vspace{5ex}
\hspace{45mm}
\mbox{\epsfxsize=40mm \epsfbox{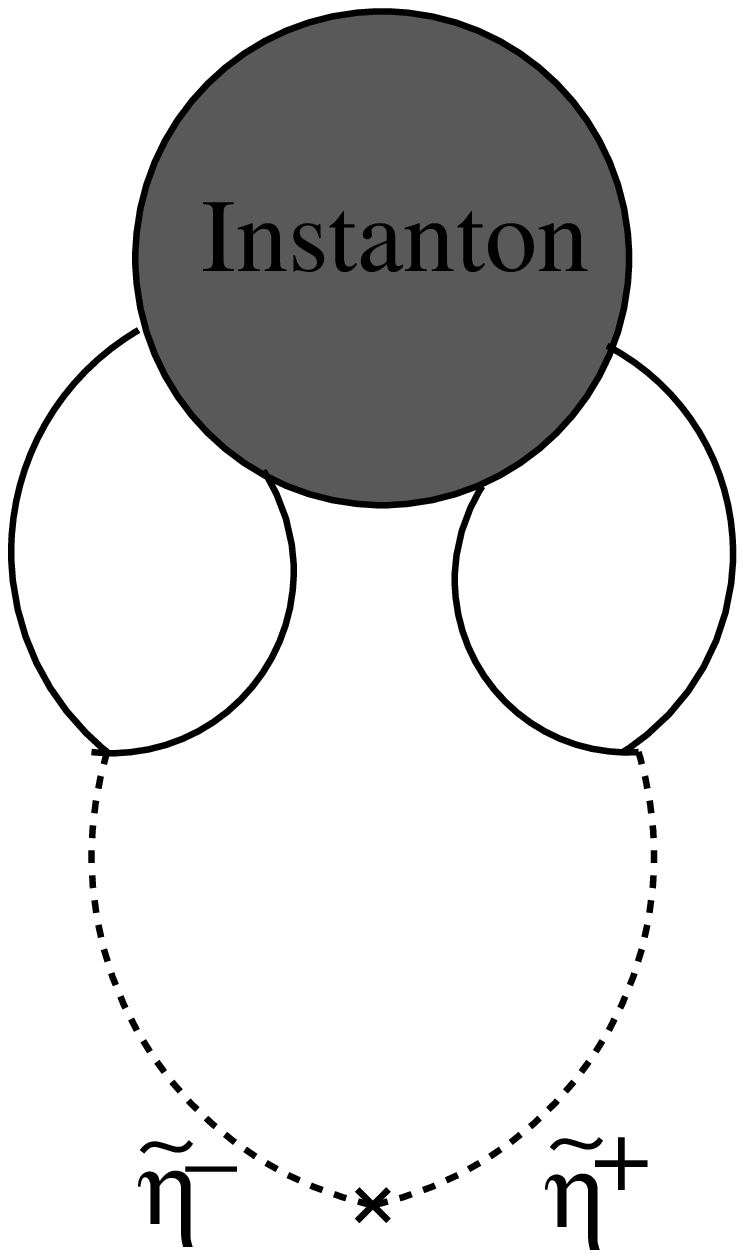}}

\begin{center}
Figure 1.  The $\svev{\tilde\h\tilde\h}$ condensate
\end{center}

\noindent The condensate that we calculate is
\beq
   \G_0 = \e_{AB} \svev{\tilde\h^+_A(0)\, \tilde\h^-_B(0)} \,.
\label{gm0}
\eeq
The relevant diagram is shown in Fig.~1.
In this diagram, each scalar line emanates  from a source which is the product
of two fermionic zero modes. The two contributions to each source, arising from
picking different channels of each zero mode, are depicted in Fig.~2.
 The condensate is given by\footnote{
  The euclidean partition function of supersymmetric theories is defined
  using a Majorana representation for the fermions~\cite{maj}. Consistency of
  this representation requires one to choose all the radial functions in
  \seeqs{gg} and\seneq{lept} to be real.}
\bqry
    \G_0 & = & c\, \e_{kl} \e_{AB}
    \int{d^4x_0 d\r \over \r^5}\, e^{-S_E(\r)}
    \int d^4z\, I_{kC}(z-x_0)\, G_{CB}(z,0;x_0)  \NON
    & &  \times
    \int d^4z'\, I_{lD}(z'-x_0)\, G_{DA}(z',0;x_0)
\eqry
where
\beq
   I_{kC} =  ig\sqrt{2}\, T^c_{CB}\, \h_{B\a}\, \e_{\a\b}\, (\l^c_\b)_k
           + y\, \e_{ij} \e_{BC} \e_{\a\b}\, \x\dg_{j\b}\, (\j\dg_{Bi\a})_k
\label{src}
\eeq
An integration over the SU(2) collective coordinates,
which yields a factor of the group's volume, has been absorbed into the
dimensionful constant $c$.

  We will now show that $\G_0$ is non-zero.
First, it is a matter of straightforward algebra to show that
\beq
   I_{kA}(x-x_0) = -i\, \e_{kA}\, s(r) \,,
\eeq

\vspace{5ex}
\hspace{20mm}
\mbox{\epsfxsize=100mm \epsfbox{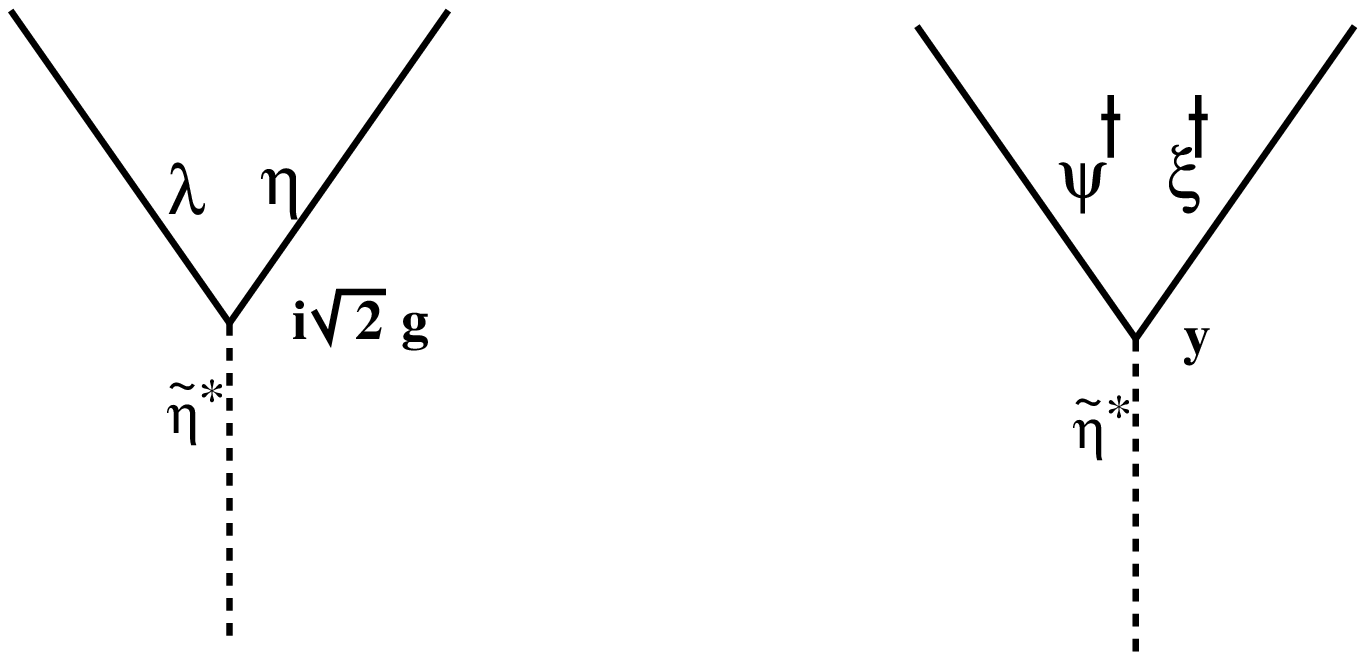}}

\begin{center}
Figure 2.  The two contributions to each source \seeq{src}
\end{center}

\noindent where
\beq
   s(r) = {3g\over \sqrt{2}} u(r) f(r) +
	  y (2 g(r) + r^2 h(r)) v(r) \,.
\eeq
In the large-$r$ limit, $s(r)$ is dominated by the $h(r)v(r)$ term.
(We assume $m<\m$, see App.~A). Hence, $s(r)$ cannot be zero everywhere.
Its asymptotic behaviour is
\beq
  s(r) \sim r^{-3} e^{-(m+m_1)r} \,.
\label{s}
\eeq
Next, the scalar propagator is defined by the equation
\beq
   \left(- D^2 + m_1^2 r^2 \varphi^2(r)\right)
   G(x,y;x_0) = \d^4(x-y) \,.
\label{prop}
\eeq
It follows from this equation that
\beq
  \e_{AB} \e_{CD}\, G_{BD}  = G^*_{AC} \,.
\eeq
Finally, we let
\beq
   F(x,x_0) = \int d^4z\, G(z,x;x_0)\, s\left( |z-x_0| \right) \,.
\eeq
Notice that $F(x,x_0)=F(x-x_0)$.
Putting everything together, the condensate is
\beq
  \G_0 = c \int{d^4x_0 d\r \over \r^5}\, e^{-S_E(\r)}\,
            \tr F(x_0) F\dg(x_0) \,,
\label{ggg}
\eeq
\seEq{s} and\seneq{prop} imply that $F(x_0)$ cannot be zero everywhere, and
that its asymptotic behaviour is the same as the $\h$-propagator, \ie
\beq
  F(x-x_0) \sim  r^{-{3\over 2}} e^{- m_1 r} \,.
\label{F}
\eeq
This complete the proof that $\G_0$ is non-zero.

  If one considers an antiinstanton instead of an instanton, one finds a
condensate of the complex conjugate fields, which satisfies the on-shell
relation $\svev{\tilde\h^* \tilde\h^*} = \svev{\tilde\h \tilde\h}^*$.
Hence, the one-instanton result \seeq{ggg} is actually the value of the
condensate
in the presence of a dilute instanton-antiinstanton gas.

\vspace{5ex}
\noindent {\large\bf 4.~~Explicit SUSY breaking}
\vspace{3ex}

 The condensate discussed in the last section can be regarded as the point-like
limit of the two-point function $\e_{AB} \svev{\h^+_A(x) \h^-_B(y)}$.
But this two-point function is not gauge invariant.
Instead, we consider the gauge invariant two-point function related by
complementarity
\beq
   \G(x,y) = \e_{AB} \e_{CD} \e_{ij}\,
             \svev{\tilde\h^+_A(x) \f_{iB}(x) \f_{jD}(y) \tilde\h^-_C(y)}\,.
\label{corr}
\eeq
The leading order contribution to $\G(x,y)$ is obtained by substituting the
classical Higgs field of \seeq{higgs} for $\f_{iA}(x)$.
In the point-like limit one obtains a new condensate $\G(x,x)=\G(0,0)$.
Computing this condensate amount to almost exactly repeating the previous
calculation. The result is
\beq
  \G(0,0) = c\, v^2 \int{d^4x_0 d\r \over \r^5}\, e^{-S_E(\r)}\,
              x_0^2\, \varphi^2(|x_0|)\,
              \tr F(x_0) F\dg(x_0) \,.
\eeq
Hence, this condensate too is non-zero.

  Now, the fields $\f_{iA}(x)$ and $\h^\pm(x)$ are all lowest components of
chiral superfields. Unbroken
SUSY requires the correlation function of any product of these fields (but
not their complex conjugates) to be a constant, independent of the separation
between points~\cite{itep,rv}.
We have seen above that $\G(x,x)$ is non-zero.
We will now prove that SUSY is explicitly broken in
the one instanton sector, by showing that $\G(x,y)$ {\it depends} on $x-y$.
In fact, $\G(x,y)\to 0$ as $|x-y|\to\infty$.

\vspace{10ex}
\hspace{20mm}
\mbox{\epsfxsize=100mm \epsfbox{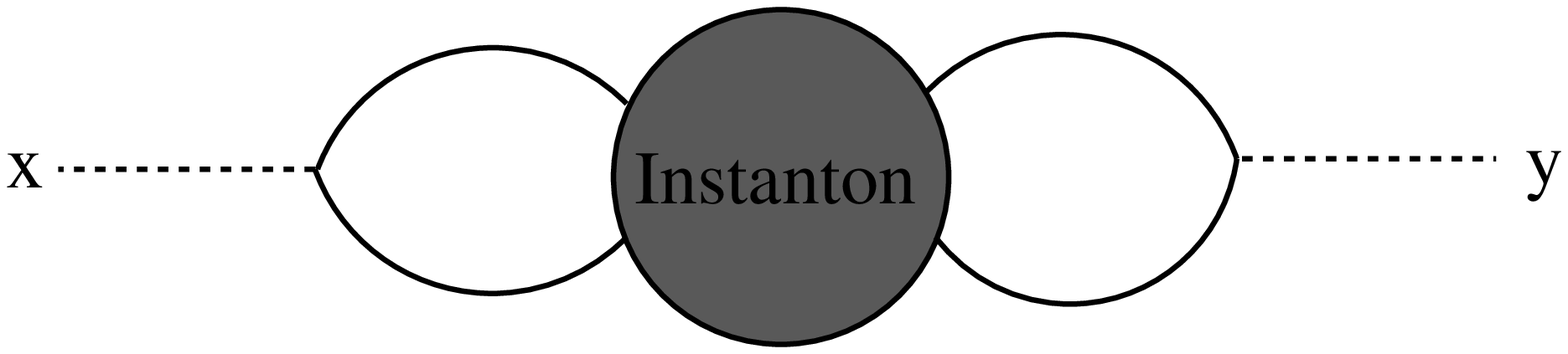}}

\vspace{3ex}
\begin{center}
Figure 3.  The two point function \seeq{corr}
\end{center}
\vspace{3ex}

\noindent $\G(x,y)$ is given by the diagram shown in Fig.~3.
Explicitly we find
\bqry
  \G(x,y) & = & c\, \int{d^4x_0 d\r \over \r^5}\, e^{-S_E(\r)} \NON
          & & \times \tr
             F(x-x_0)\, \f\dg(x-x_0)\,
             \f(y-x_0)\, F\dg(y-x_0) \,.
\eqry
Using \seeqs{higgs} and\seneq{F} the asymptotic behaviour is
\beq
  \G(x,y) \sim |x-y|^{-{1\over 2}} e^{- m_1 |x-y|} \,.
\eeq
Thus, $\G(x,y)$ tends to zero exponentially at large separations.

\vspace{5ex}
\noindent {\large\bf 5.~~Inducing the Electro-Weak scale}
\vspace{3ex}

  In this section we illustrate a new mechanism for
inducing the Electro-Weak scale via non-perturbative SUSY
breaking effects. To this end, consider the following superpotential
\beq
  W' =  h' H^0 \left(
       {1\over 2} \e_{i'j'} \e_{A'B'} H_{i'A'} H_{j'B'} -
       \e_{AB} \h^+_A \h^-_B
       \right)\,.
\label{ew}
\eeq
The $H_{i'A'}$ are now assumed to be the two Higgs superfields of the minimal
supersymmetric Standard Model, and the primed indexes refer to
SU(2)$_L$. The model introduced in Sect.~2 is now regarded as a prototype
for some higher scale physics, \ie as a hidden sector.

  The scalar potential that corresponds to \seeq{ew} contains a term
\beq
  V =  h'^2  \left|
       {1\over 2} \e_{i'j'} \e_{A'B'} \tilde{H}_{i'A'} \tilde{H}_{j'B'} -
       \e_{AB} \tilde\h^+_A \tilde\h^-_B
       \right|^2\,.
\label{v}
\eeq
Now, once the condensate $\svev{\tilde\h \tilde\h}$ is formed,
this becomes (see \seeq{gm0})
\beq
  V =  h'^2  \left|
       {1\over 2} \e_{i'j'} \e_{A'B'} \tilde{H}_{i'A'} \tilde{H}_{j'B'} -  \G_0
       \right|^2\,.
\label{vv}
\eeq
As a result, the Higgs fields of the supersymmetric Standard Model develop
an expectation value  $\svev{\tilde{H}_{i'A'}} = \G_0^{\half} \d_{i'A'}$, which
breaks
SU(2)$\times$U(1)$_Y$ to U(1)$_{EM}$. Notice that going from \seeq{v} to
\seeq{vv} leaves the lagrangian of the observed sector supersymmetric. We do
expect,
however, that the inclusion of other effective interactions induced by
the hidden sector will give rise to explicit SUSY breaking terms in the
lagrangian
of the observed sector.

  Can this mechanism yield a phenomenologically
acceptable value for the Electro-Weak scale?
There are two possible sources for $\G_0$. One scenario
is that the condensate arises from GUT scale physics. In this
case, the model of Sect.~2 should be regarded as a toy model for
the relevant GUT scale physics.

  The other scenario is based on the observation that
the non-perturbative effects violate the SUSY algebra.
Consequently, {\it negative} values for the vacuum energy are not
impossible. One should investigate the possibility that, as
in ordinary QCD, the non-perturbative effects in supersymmetric QCD {\it lower}
the vacuum energy~\cite{s}. If this is true, supersymmetric QCD will exist in a
confining phase where the SUSY violating effects are $O(1)$.
The Electro-Weak scale can then be induced by a strongly interacting
hidden sector at the TeV range.

\vspace{5ex}
\noindent {\large\bf 6.~~Discussion}
\vspace{3ex}

  In this paper we showed that one instanton effects in SUSY-Higgs models
violate SUSY explicitly.
How does this result compare with previous calculations?
Non-perturbative SUSY breaking effects have already been found in
ref.~\cite{cs1,cs2}. We should mention in particular the demonstration that
the $S$-matrix for elementary particle -- soliton scattering is not
supersymmetric already at tree level~\cite{cs2}.

  In the literature on supersymmetric Yang-Mills (SYM) there are
one instanton calculations that give rise to supersymmetric results (see \eg
ref.~\cite{it,itep}). But SYM is a strongly interacting theory, and so the
one instanton result in SYM is not a leading order
contribution in any systematic expansion. For example, in the case
of an SU(2) theory, the correlator $\svev{\l\l(x)\l\l(y)}$ is non-zero
on the one hand, and it is required to be independent of the separation
$x-y$ on the other hand~\cite{itep,arv}. The one instanton contribution to this
correlator is dominated by instantons whose size $\r$
satisfies $\r\sim|x-y|$. Hence,
the supersymmetric result is unreliable for separations which are large
compared to the confinement scale. A similar  statement applies to
instanton calculations in supersymmetric QCD.
Since the squarks' VEV can potentially be zero, one cannot rule out the
possibility that the theory is strongly interacting {\it and} breaks SUSY
explicitly at the same time.

  The existence of explicit non-perturbative SUSY breaking effects raises some
as yet unresolved issues. In the case of the chiral anomaly, the local
continuity equation is violated in perturbation theory. This entails
a violation of the axial charge at the non-perturbative level, whose
manifestation is the occurrence of fermionic zero modes~\cite{tft}.
Since we have found that conservation of the SUSY charge is violated by
non-perturbative effects, the question arises whether there is some indication
from perturbation theory that this is going to happen.

  Present day understanding of the perturbative properties of
the SUSY current leaves open certain subtleties.
At the moment, we would like to draw attention to some
general differences between axial symmetries and SUSY.
The chiral anomaly is a phenomenon that occurs at the level of a free
fermion field in an external gauge field. In this setting perturbation theory
has a finite radius of convergence. In fact, a fermionic determinant whose
gauge variation is given exactly by the usual anomaly can be defined for
non-perturbative gauge fields as well~\cite{tft1}.

  In the SUSY case, on the other hand, it is impossible to consider the
gauge field  as external without breaking the supermultiplet structure.
Because of the non-linearity of the SUSY current, the definition of a conserved
current can only be done order by order in perturbation theory~\cite{pert}.
In a full-fledged field theory, however,
perturbation theory is only an asymptotic expansion, whose minimal error is
given by the magnitude of non-perturbative effects.
Thus, the perturbative construction only implies
that violations of the conservation equation, if they exist,
must be of a non-perturbative nature.

\vspace{5ex}
\noindent {\large\bf Appendix A. Asymptotic behaviour of the zero modes}
\vspace{3ex}

  In this appendix we discuss the asymptotic behaviour of the zero modes.
The ``lepton'' zero modes solve the following set of ordinary differential
equations
\bqry
  2u' + 3 a\, u + m_1 \varphi\, v & = & 0 \,, \NON
  2v' + m_1 \varphi\, u & = & 0 \,.
\eqry
Here $a=a(r)$ and $\varphi=\varphi(r)$, see \seeqs{a} and\seneq{higgs}.
The prime denotes differentiation with respect to $r^2$.
The asymptotic large-$r$ behaviour inferred from these equations is
\beq
  u(r)\,, v(r) \sim r^{-{3\over 2}} e^{-m_1 r} \,.
\eeq

  In order to write down the equations for the gaugino zero modes
we introduce the linear combination
\beq
   h_1(r) = g(r) + 2 r^2 h(r) \,.
\eeq
The radial equations are
\bqry
  2f' + 4 a f + (\m/\sqrt{2}) \varphi\, g & = & 0 \,, \NON
  2g' + (r^{-2}-2a)g + (a-r^{-2})h_1 +\sqrt{2}\m\varphi f & = & 0 \,, \NON
  2p' - (m/\sqrt{2}) \varphi\, h_1 & = & 0 \,, \NON
  2h'_1 + (3/r^2)h_1 + 3(a-r^{-2})g - \sqrt{2} m\varphi\, p & = & 0 \,.
\eqry
The pairs $(f,g)$ and $(p,h_1)$ diagonalize the mass operator at
infinity. Notice also that the mixed terms in the $g$- and $h_1$-equations
are proportional to $a(r)-r^{-2}$, which decreases exponentially for large $r$.
It will be convenient for us to consider the case $m<\m$.
The asymptotic behaviour of each channel is then determined by its own mass,
and we find
\bqry
  h_1(r)\,, p(r) & \sim & r^{-{3\over 2 }} e^{-m r} \,, \\
  f(r)\,, g(r)   & \sim & r^{-{3\over 2 }} e^{-\m r}\,.
\eqry

\newpage
\noindent {\large\bf Appendix B. Generalization to $N_f > 2$}
\vspace{3ex}

  In this appendix we show that the same pattern found for $N_f=2$
generalizes to $N_f > 2$. Again, we will show that a correlation
function which is required by SUSY to be a constant, fails in fact to be so.

  The model with  $N_f > 2$ is constructed as follows. Instead of two
lepton families we now take $2M$ families where $M=N_f-1$. The corresponding
superfields are denoted $\h_A^{n\pm}$ and  $\x_i^{n\pm}$, where
$n=1,\ldots,M$. We also introduce two new neutral superfields $\o_1$ and
$\o_2$.
These will form a massive Dirac fermion and two massive scalars which are
singlets under all the internal symmetries except the non-anomalous
$R$-symmetry. The role of the new scalars is to absorb the additional zero
modes present for $N_f > 2$.

  The superpotential is
\beq
  W = W_1 + \sum_{n=1}^M W_2(\h_A^{n\pm},\x_i^{n\pm}) + W_3 \,,
\eeq
where $W_1$ and $W_2$ are given by \seeqs{sp} and\seneq{sp2} respectively, and
\beq
  W_3 = m' \o_1 \o_2\,
        +\, y' \e_{ij}\,\o_1 \sum_{n=1}^M  \x_i^{n+} \x_j^{n-}  \,.
\eeq
The general model, too, has a unique supersymmetric minimum,
and the VEV-s of all the new scalar fields are zero.
As in the $N_f=2$ case, all fields acquire non-zero masses at tree level.

\vspace{10ex}
\hspace{20mm}
\mbox{\epsfxsize=100mm \epsfbox{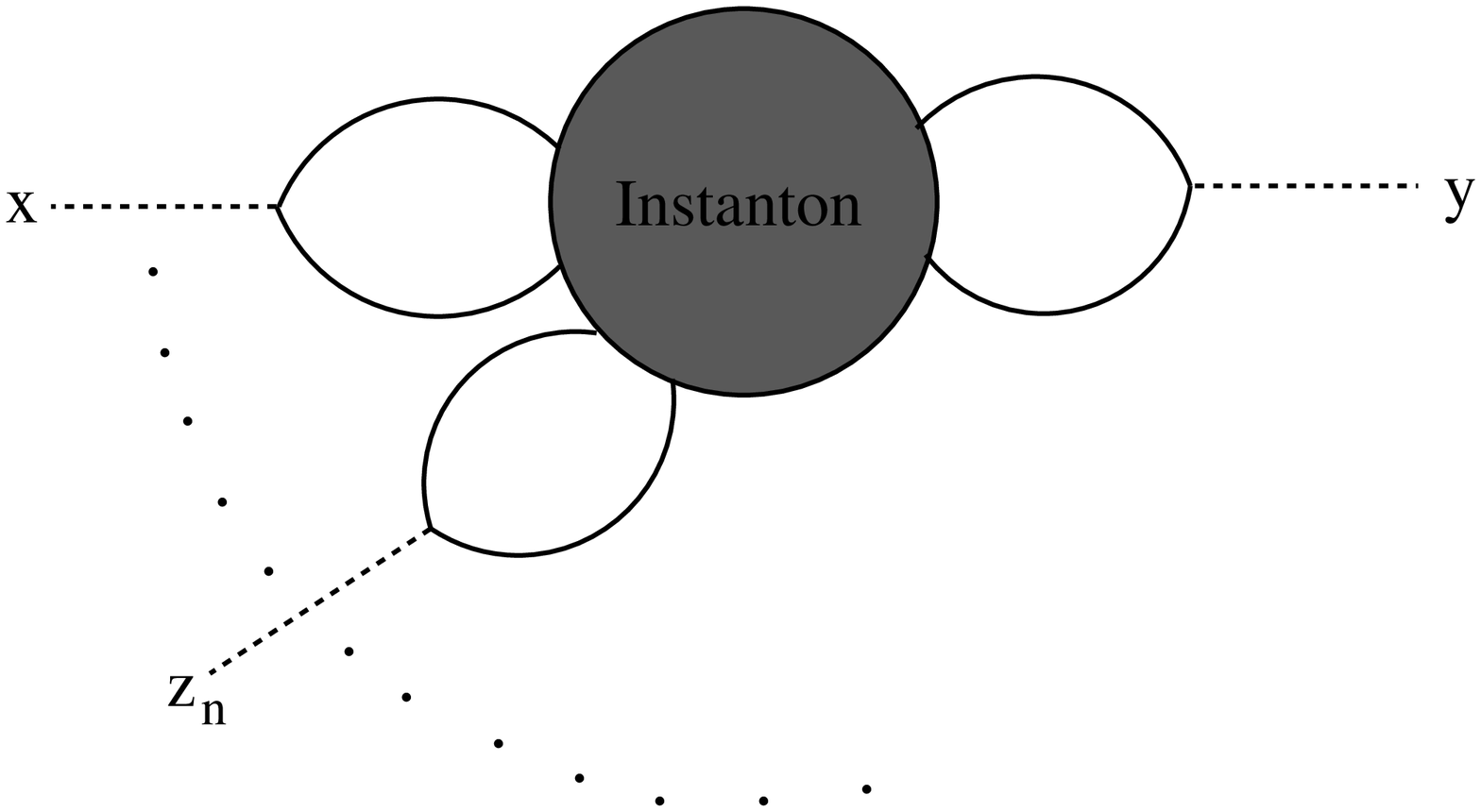}}

\vspace{5ex}
\begin{center}
Figure 4.  The correlation function \seeq{gz}
\end{center}

\newpage
  The gauge invariant correlator that unbroken SUSY would require to be
a constant is (compare \seeq{corr})
\bqry
   \G(x,y,z_1,\ldots,z_{M-1}) & = & \e_{AB} \e_{CD} \e_{ij}\, \NON
   & & \times \svev{\tilde\h^+_A(x) \f_{iB}(x) \f_{jD}(y) \tilde\h^-_C(y)
   \tilde\o_1(z_1)\cdots \tilde\o_1(z_{M-1})}\,.
\label{gz}
\eqry

\vspace{10ex}
\hspace{40mm}
\mbox{\epsfxsize=40mm \epsfbox{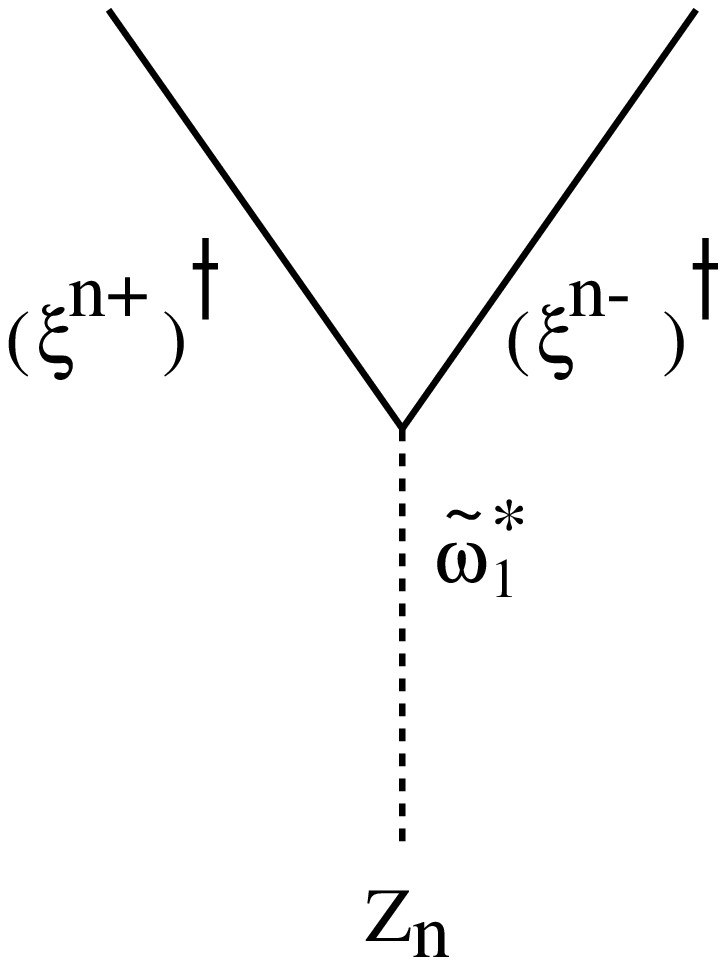}}

\begin{center}
Figure 5.  The source \seeq{t}
\end{center}
\vspace{5ex}

\noindent One finds (see figs.~4 and 5)
\bqry
  \G(x,y,z_1,\ldots,z_{M-1}) & = & c M!\,
             \int{d^4x_0 d\r \over \r^5}\, e^{-S_E(\r)} \NON
          & & \times \tr
             F(x-x_0)\, \f\dg(x-x_0)\,
             \f(y-x_0)\, F\dg(y-x_0) \NON
          & & \times
             T(z_1-x_0)\cdots T(z_{M-1}-x_0)\,,
\eqry
where
\beq
  T(z-x_0) = 2y' \int d^4x\, v^2(|x-x_0|)\, G_1(x,z;x_0) \,,
\label{t}
\eeq
and $G_1$ is the $\tilde\o$-propagator. As in Sect.~4, one can
show that in the point-like limit $x=y=z_1=\ldots=z_{M-1}$
one obtains a non-zero condensate, whereas $\G(x,y,z_1,\ldots,z_{M-1})$ tends
to zero if the separation between any two points tends to infinity.

\vspace{5ex}
\centerline{\rule{5cm}{.3mm}}



\newpage
%
%
\begin{table}[h]
\begin{center}
\begin{tabular}{|c|c|c|c|}     \hline
 superfield     & fermion & boson & $Q_R$ \\ \hline
$V^a$ & $\l^a$ & $A^a_\m$  & 1 \\ \hline
$\Phi_{iA}$  & $\psi_{iA}$  & $\phi_{iA}$ &  -1 \\ \hline
$\Phi^{0}$   & $\psi^{0}$   & $\phi^{0}$ & 1\\ \hline
$\h^\pm_A$ & $\h^\pm_A$ & $\tilde\h^\pm_A$ & -1 \\ \hline
$\x^\pm_i$    & $\x^\pm_i$    & $\tilde\x^\pm_i$ &  1 \\ \hline
\end{tabular}
\caption{The field content of the model of Sect.~2. The last row gives
the fermion's charge under the non-anomalous $R$-symmetry.}
\vspace{1.0cm}
\end{center}
\end{table}

%
%
\begin{table}[h]
\begin{center}
\begin{tabular}{|c|c c c c c c c|}     \hline
channel &
$S_{1}$ &
$S_{2}$ &
 T      &
 F      &
 L      &
$K_{1}$ &
$K_{2}$   \\ \hline \hline
$\l^a_\a$ &
$\frac{1}{2}$ & 0 & 1 & 0 & 0 &
$\frac{1}{2}$ & 0  \\ \hline
$\j\dg_{iA\a}$ & 0 & $\frac{1}{2}$ &
$\frac{1}{2}$ &
$\frac{1}{2}$ & 0,1 &
$\frac{1}{2}$ &
0  \\ \hline
$\j^0_\a$ & $\frac{1}{2}$ & 0 & 0 & 0 & 0 &
$\frac{1}{2}$ &
 0 \\ \hline \hline
$\h_{\a A}$  & $\frac{1}{2}$ & 0 & $\frac{1}{2}$ & 0 & 0 &
0 & 0 \\ \hline
$\x\dg_{\a i}$  & 0 & $\frac{1}{2}$ & 0 & $\frac{1}{2}$ &
0 & 0 & 0 \\ \hline
\end{tabular}
\caption{The channels of the fermionic zero modes and their quantum
numbers.}
\end{center}
\end{table}

\end{document}